\documentclass[conference]{IEEEtran}
\IEEEoverridecommandlockouts
\usepackage{cite}
\usepackage{amsmath,amssymb,amsfonts}
\usepackage{algorithmic}
\usepackage[ruled,vlined]{algorithm2e}
\usepackage{graphicx}
\usepackage{epstopdf}
\usepackage{caption}
\usepackage{subcaption}
\usepackage{textcomp}
\usepackage{xcolor}
\def\BibTeX{{\rm B\kern-.05em{\sc i\kern-.025em b}\kern-.08em
    T\kern-.1667em\lower.7ex\hbox{E}\kern-.125emX}}
\begin{document}

\title{Priority-Queue based Dynamic Scaling for Efficient Resource Allocation in Fog Computing\\
% {\footnotesize \textsuperscript{*}Note: Sub-titles are not captured in Xplore and
% should not be used}
% \thanks{Identify applicable funding agency here. If none, delete this.}
 }

\author{
\IEEEauthorblockN{Saksham Bhushan}
\IEEEauthorblockA{\textit{Department of EECS} \\
\textit{Indian Institute of Technology Bhilai}\\
Raipur, India \\
sakshamb@iitbhilai.ac.in}
\and
\IEEEauthorblockN{Maode Ma}
\IEEEauthorblockA{\textit{School of Electrical \& Electronic Engineering} \\
\textit{Nanyang Technological University}\\
Singapore \\
EMDMa@ntu.edu.sg}
% \and
% \IEEEauthorblockN{3\textsuperscript{rd} Given Name Surname}
% \IEEEauthorblockA{\textit{dept. name of organization (of Aff.)} \\
% \textit{name of organization (of Aff.)}\\
% City, Country \\
% email address or ORCID}
% \and
% \IEEEauthorblockN{4\textsuperscript{th} Given Name Surname}
% \IEEEauthorblockA{\textit{dept. name of organization (of Aff.)} \\
% \textit{name of organization (of Aff.)}\\
% City, Country \\
% email address or ORCID}
% \and
% \IEEEauthorblockN{5\textsuperscript{th} Given Name Surname}
% \IEEEauthorblockA{\textit{dept. name of organization (of Aff.)} \\
% \textit{name of organization (of Aff.)}\\
% City, Country \\
% email address or ORCID}
% \and
% \IEEEauthorblockN{6\textsuperscript{th} Given Name Surname}
% \IEEEauthorblockA{\textit{dept. name of organization (of Aff.)} \\
% \textit{name of organization (of Aff.)}\\
% City, Country \\
% email address or ORCID}
}

\maketitle

\begin{abstract}
In this emerging world of connected devices, the need for more computing devices with a focus on delay-sensitive application is critical. In this paper, we propose a priority-queue based Fog computing architecture combined with dynamic scalability of fog devices, which not only reduces the delay experienced by delay-sensitive tasks by categorizing the delay-sensitive and delay-insensitive tasks, but also dynamically allocates the fog devices within the network depending upon the computation load for reducing the power consumption. The results show that the proposed algorithm is able to achieve a significant lower delay for both delay-sensitive and -insensitive tasks when compared with other related schemes with a $14.5\%$ lower power consumption.
\end{abstract}


\section{Introduction}
With the increasing accessibility to cloud computing infrastructure, the demand for cloud services has drastically increased. Cloud services, these days, are used for off-device computing and storage. It has also bolstered the growth of connected devices such as Internet of Things (IoT). IoT is considered to be a crucial part of the Industry 4.0 revolution. IoT devices are physical devices which are responsible for collecting and sharing data for various use cases such as smart homes, smart cities, e-health care, etc. The number of IoT devices are expected to raise to 43 billion in 2023, a threefold increase from 2018 \cite{mckinsey}. Major drivers of this widespread adaptation of IoT is due to the low costs of sensors, and edge computing devices, penetration of cloud services and mobile computing devices \cite{delloite}.

IoT devices rely on computation and storage by the cloud computers. The immense amount of data collected by IoT devices are sent to cloud data centres where it is processed and stored and in some cases the cloud responds with the necessary actions required back to the IoT device. The cloud data centres are generally distant from the IoT devices and handling such huge influx of data creates a network bottleneck at the data centre, also induces latency and low Quality of Service (QoS) for the IoT service. Also, for responding to such huge amount of requests from various connected IoT devices the data centres will have to run round the clock at almost same potential. Most of the requests from the IoT devices require trivial computations and it is inefficient to ping the data centre every time for these requests. With the ever increasing number of IoT devices connected to the cloud, this problem will become more predominant. This is where the concept of Fog Computing comes in.

The term Fog computing was coined by Cisco in 2012 and it is a distributed computing approach which offloads computation from the cloud to the edge devices, hence From-clOud-to-eGde (FOG) \cite{cisco}.  It enables computation and storage at network devices at different hierarchical levels with different computation and storage capacity. Since the request handling is done in close proximity to the IoT device, it allows for a better QoS to the end user by reducing the delay for time sensitive applications and enabling efficient use of resources. Thus the major challenges for Fog environment for IoT applications are task scheduling and resource allocation \cite{survey2018}. Fog computing is not a replacement to cloud computing rather the two technologies complement each other. IoT devices can send trivial computations to Fog computing devices for computation and storage and Fog devices can forward the final output and the accumulated data to the cloud at once, hence reducing the number of requests to the cloud. For sophisticated computations IoT devices can still communicate directly with the cloud services.

For proper and efficient coordination between Fog devices and cloud, the computation tasks are needed to be properly allocated between the cloud and Fog nodes. Thus, to analyse such a resource allocation method, we use queuing theory for the performance analysis. Queuing theory has been extensively used for such analysis as it provides insights on various QoS factors such as system response time, CPU utilization, mean throughput, etc. 
The aim of Fog computing is to provide better QoS for delay-sensitive applications but due to the limited computation available on the Fog devices it is not possible to complete every allocated task within a delay-threshold. Thus, in this paper, we propose a priority service provision scheme where tasks can be classified into delay-sensitive and delay-insensitive tasks. The delay-sensitive and delay-insensitive tasks can be defined as high priority and low priority tasks respectively. 
The contribution of this paper can be summarized in these points:
\begin{itemize}
    \item We provide an queuing analytical model for priority based task scheduling in Fog-cloud architecture.
    \item We provide a model for dynamic scalability of Fog nodes for energy efficient resource allocation.
    \item We evaluate our proposed model using the analytical model and compare with other existing schemes and verify the results with JMT simulator.
\end{itemize}

The rest of the paper is organized in the following manner:  Section \ref{related} summarizes the related works. Section \ref{sysarch} explains the fog computing architecture. Section \ref{promodel} illustrates the proposed priority service provision and the dynamic scaling of fog nodes. Results and simulation setup are demonstrated in Section \ref{results}. Finally, Section \ref{conclusion} provides concluding remarks and point the future works.

\section{Related Works}\label{related}
Resource allocation and task scheduling for edge-fog-cloud architectures have been a subject of thorough research in recent years. S. Misra \emph{et al.} in \cite{multiarmed} propose an approach where high-level tasks are divided into smaller independent subtasks and distributed among the Fog nodes using a greedy approach. In \cite{detour,decentral} authors propose greedy and game theoretic approaches, respectively, for deciding whether to offload the tasks to nearby fog nodes or cloud services by considering factors such as delays, energy consumption, etc. \cite{intelli} models the fog computing environment as a Markov discrete process, where dynamic fog node mobility and resources availability are considered and then proposes an online resource allocation algorithm aiming at maximizing the number of satisfied user requests within a predefined delay threshold. In \cite{sfc}, authors propose a network function virtualization based service function chain (SFC) provisioning from cloud networks to the fog environments, it shows a first look at SFC provisioning in a multi-layer fog architecture that considers client resource and delay requirements. In \cite{sfcpriority}, the authors then extend the work by proposing a priority based SFC scheme for efficient resource allocation.

Many researches aim towards performance analysis of Fog computing architectures using queuing theory. In \cite{performance}, authors propose a queuing theory based framework and edge-fog-cloud architecture for performance evaluation in IoT application. FogQN \cite{fogqn} provides an analytical model and a tool based on open multi-class Queuing Networks (QN) for fog and cloud computing. The paper aims to allocate tasks between cloud and fog servers by underpinning optimal fraction $f$ of data processing executed. In \cite{dynapriority}, authors propose a priority based task scheduling to reduce the response time for delay-sensitive tasks and then propose a dynamic priority scheme for non-computing tasks to maintain the delay within a certain threshold.

For evaluating the effect of scalability on the fog network, authors in \cite{effanddyna} propose a queuing theory based model for efficient scaling of the fog nodes such as to satisfy the QoS requirements of IoT application. In \cite{scalable}, the authors propose a scalable design and dimensioning of a fog infrastructure via a mixed-integer linear program to construct a physical fog network design by mapping IoT virtual networks to dimensioned fog nodes.

Though there have been many proposals for efficient task scheduling and resource allocation in fog computing paradigm but, to the best of our knowledge, the proposals present in the literature are fairly complex and provides separate approaches for efficient resource allocation and task scheduling. Hence, tackling both issues using two separate models at once will further increase the complexity. Therefore, we think that the literature strictly needs an amalgamated solution for both the major problems in fog computing paradigm. Hence, we propose a simple and effective scalability model which tackles the issue of both resource allocation and task scheduling at once. We also justify our study by providing thorough study of the power consumption by the scalable fog nodes.

\section{Fog computing system architecture}\label{sysarch}

\begin{figure}[h]
\includegraphics[width=5cm]{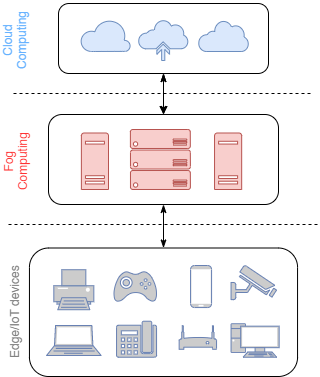}
\centering
\caption{Architecture of edge-fog-cloud system.}
\label{fig:arch}
\end{figure}

The basic architecture of the edge-fog-cloud architecture can be divided into three layers laid down in a hierarchical manner \cite{Naha_2019}. Fig. \ref{fig:arch} shows the high-level hierarchical structure of the edge-fog-cloud architecture. The layers of a edge-fog-cloud architecture can be divided into three layers based upon the main actor of the Fog environment, namely, cloud computing layer (CCL), fog computing layer (FCL) and edge/IoT layer (EL).

In the conventional edge-cloud model, the edge devices directly communicate with the CCL and the CCL is responsible for processing and storage of the data accumulated from the EL. But this direct communication between CCL and EL induces huge delay and unnecessary load on the CCL. Hence, FCL is introduced as an intermediate layer for catering delay-sensitive applications. The computation is done only by CCL and FCL.
\subsection{Fog computing layer}
FCL is responsible for data collection from the edge devices and then deciding whether the data is required to be sent to the CCL for further processing. Small and trivial processing can be done at the FCL itself and the final results can be shared with the edge device for further actions and with the CCL for storage. For delay-sensitive applications, this layer is the most important layer, as the fog nodes in this layer are within close proximity of the edge devices. The fog nodes can be a dedicated devices for computation or a certain amount of computation power can be dedicated on-demand for fog computation from already existing devices in the network.
\subsection{Cloud computing layer}
CCL consists of the cloud infrastructure, the function of this layer is long-term data storage and processing of complex time-insensitive computations. This layer relies totally on FCL for receiving data from the edge devices. This layer is complementary to the FCL as FCL has very limited computation and storage capabilities which makes CCL crucial and CCL is not in close proximity of the edge devices, hence it induces significant delay, thus time-sensitive applications cannot only rely on CCL.
\subsection{Edge/IoT layer}
This is the data generation layer, consisting of all the IoT or edge devices in the network. These devices are often equipped with sensors or actuators which collect data which is to be forwarded to the FCL for further processing. These devices range from everyday applications such as smart home devices, health monitoring devices, entertainment systems, etc. to critical applications such as alarm systems, security systems, sensors as in chemical labs and factories and many more. These devices are connect to FCL through a two-way communication channel, which lets FCL respond back to the devices within a time threshold after processing the tasks natively.

For the communication between edge, fog and cloud layers, a separate data channel needs to be implemented which will carry information about the state of the whole network, resource and tasks allocation information. These tasks are critical as they are required to be processed by respective layers for the information retrieval, fault detection and error logging. Thus, we propose to call these tasks System critical tasks (SCT). These tasks are required to be communicated and processed at certain interval, thus they must have a defined threshold response time. To make things simpler we can consider them to be a part of the regular delay-sensitive and delay-insensitive classes and based upon the respective response time we can dynamically allocate these tasks in either of the classes.

\section{Proposed model}\label{promodel}

\begin{figure*}[t]
\includegraphics[width=0.8\textwidth]{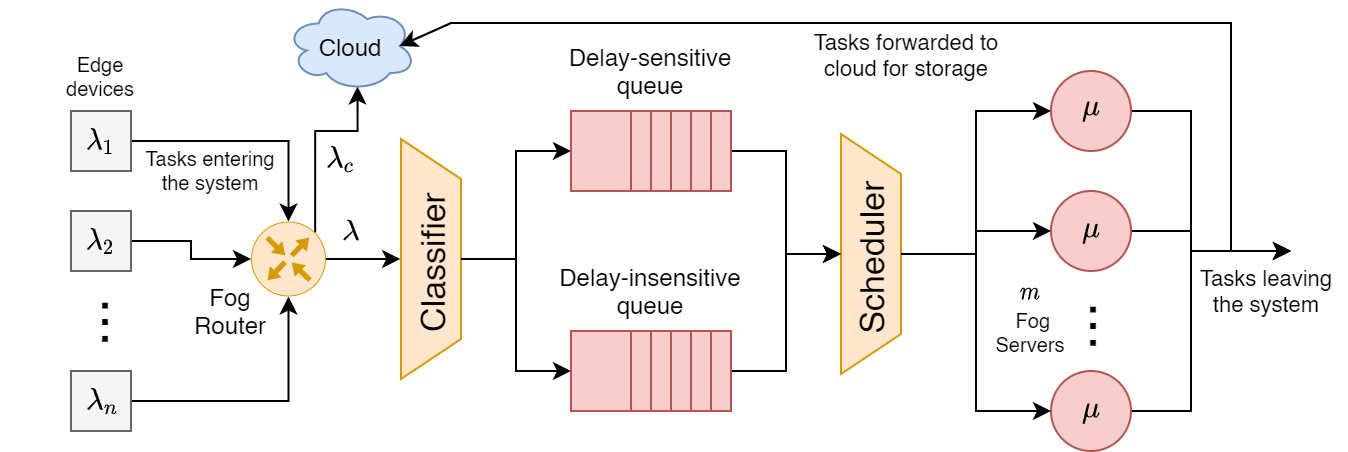}
\centering
\caption{Queueing model of the Fog computing system.}
\label{fig:queue}
\end{figure*}

The Fog computing model follows an M/M/m queue, where m denotes the number of fog nodes or servers in the system. Detailed structure of the queueing model for the fog computing system is shown in Fig. \ref{fig:queue}. The tasks in the system enter from the edge/IoT layer. We have assumed $n$ number of edge devices sending data or computing tasks where the rate of task arrival from $i^{th}$ devices is given by $\lambda_i$. The tasks are passed through a fog router which decides whether the requested tasks is to be forwarded to the CCL or is to handled by FCL. Therefore, we assume the rate of tasks forwarded to CCL without computation at FCL be $\lambda_c$. Hence, we get the net rate of task arrival in the FCL as,
\begin{equation}
    \lambda = \sum_{i=1}^n \lambda_i - \lambda_c
\end{equation}
Therefore, for a queueing system with arrival rate $\lambda$ and service rate $\mu$ for each server, the steady state probability is given in \cite{queuebook} as
\begin{equation}
    \pi_0 = \Bigg[\sum_{k=0}^{m-1} \frac{(m\rho)^k}{k!} + \frac{(m\rho)^m}{m!}\frac{1}{1-\rho} \Bigg]^{-1}
\end{equation}
where, $\rho$ is the individual server utilization, given by $\rho=\lambda/(m\mu)$ and the condition for the queueing system to be stable is $\rho<1$.

\begin{figure}[h]
\includegraphics[width=\linewidth]{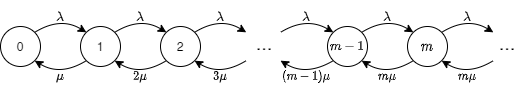}
\centering
\caption{Continuous time Markov chain for a new request within FCL.}
\label{fig:ctmc}
\end{figure}

 The continuous time Markov chain (CTMC) for new task in the FCL with respective parameters is shown in Fig. \ref{fig:ctmc}. The steady-state probability that an arriving tasks will have to wait in the queue is given by 
\begin{equation}
    P_m = \frac{(m\rho)^m}{m!(1-\rho)} \pi_0
\end{equation}
Also, mean number of tasks in the system can be given by 
\begin{equation}
    \overline{K} = m\rho + \frac{\rho}{1-\rho}P_m
\end{equation}

\subsection{Priority based Service Provision}
For implementing the priority based service provision, we classify the tasks into 2 classes:
\begin{enumerate}
    \item Class 1 - Higher priority, delay-sensitive computing tasks.
    \item Class 2 - Lower priority, delay-insensitive computing tasks.
\end{enumerate}

The tasks are assigned their respective classes by a classifier, which classifies the tasks as delay-sensitive and delay-insensitive tasks. The critical application tasks such as e-health, sensor monitoring in laboratories, factories, etc. are classified as delay-sensitive tasks whereas tasks related to mundane data logging such as temperature logging in smart homes, are classified as delay-insensitive tasks.
formulas,

Here, we consider the case where a task already in service is not preempted by an arriving task with higher priority. 
The mean remaining service time $\bar{W_0}$ of the task in service for two priority classes is given by 
\begin{equation}
    \overline{W_0} = \frac{P_m}{m\rho}\bigg(\frac{\rho_1}{\mu_1}+\frac{\rho_2}{\mu_2}\bigg)
\end{equation}
where $\rho_i$ and $\mu_i$ is the server utilization and the service rate for the tasks of class $i$, respectively.

Therefore, we can write the mean queue delay for respective classes to be 
\begin{equation}
    \overline{W_1} = \frac{\overline{W_0}}{(1-\rho_1)}
\end{equation}
\begin{equation}
    \overline{W_2} = \frac{\overline{W_0}}{(1-\rho_1-\rho_2)(1-\rho_1)}
\end{equation}

For calculating $\rho_1$ and $\rho_2$, we assume a parameter $\alpha$ which denotes the average ratio of number of high priority tasks to the number of total tasks. By using this we get 
\begin{equation}
    \rho_1 = \alpha\rho \text{ and } \rho_2 = (1-\alpha)\rho
\end{equation}
Now, lets define the threshold waiting time for SCT to be $W^{th}_{SCT}$. From \cite{dynapriority}, the waiting time for class1 and class 2 tasks are $W_1$ and $W_2$, respectively. If $W^{th}_{SCT} > W_2$, then the SCT will be counted along with delay-insensitive tasks and the waiting time for the SCT will be $\overline{W_2}$ whereas if $W^{th}_{SCT} < \overline{W_2}$, then the SCT will be counted along with delay-sensitive tasks and the waiting time for the SCT will be changed to $\overline{W_1}$. We assume the ratio of such tasks to the total number of tasks to be $\beta$. Therefore, the server utilization for class 1 and class 2, when SCT is considered in class 1 will respectively be
\begin{equation}
    \rho_1 = (\beta + \alpha)\rho \text{ and } \rho_2 = (1 - \alpha - \beta)\rho
    \label{newrho}
\end{equation}

\subsection{Provisioning Scalable Fog Nodes}

We propose a model for dynamic scaling of fog nodes based on the influx of the tasks. By this we will be able to dynamically allocate the fog nodes in FCL, for example if the number of incoming tasks is increasing then the number of fog nodes will also scale accordingly to meet the increasing demand of computation. Similarly, if the number of tasks reduces, the number of allocated fog nods will also reduce. This will allow for a more efficient usage of resources and since we propose that the fog nodes can be a part of the regular use devices within a network, this on-demand allocation of fog nodes will also be efficient and beneficial for it. For this dynamic allocation, we define a maximum and minimum number of allowed fog nodes, by this we avoid network failure as incase of increased computation tasks and some cases of network congestion, not constraining the fog allocation can have a catastrophic affect on the network. Therefore, we define the maximum and minimum allowed fog nodes as $m_{max}$ and $m_{min}$ and the waiting time of class 1 with $m$ fog servers as $\overline{W_1}(m)$. Also, we define the threshold waiting time of the class 1 tasks as $W_1^{th}$.

\begin{algorithm}[h]
\SetAlgoLined
% \KwResult{Write here the result }
 \textbf{Initialize parameters}\;
 $\lambda, \mu, m, m_{max}, m_{min}, W_1^{th}, W_{SCT}^{th}, \alpha, \beta$\\
 %$k_{add} = 0$\;
 Calculate $\rho_1$, $\rho_2$ and $\rho$\;
 \While{$\lambda < \mu m_{min}$}{
  Calculate $\overline{W_1}(m)$ and $\overline{W_2}$\;  
  \eIf{$W^{th}_{1} < \overline{W_1}(m)$}{
    %\tcp{Increase the number of fog nodes}
   \While{$W^{th}_{1} < \overline{W_1}(m)$ and $m<m_{max}$}{
   %\tcp{Allocate a new Fog node}
   Allocate a new Fog node to FCL\;
   \eIf{available fog nodes == 0}{$m_{max}=m$\;}{$m=m+1$\;}
   Calculate $\overline{W_1}(m)$\;}
   }{
   %\tcp{Decrease the number fog nodes}
   \While{$W^{th}_{1} > \overline{W_1}(m-1)$ and $m>m_{min}$}{
   %\tcp{Remove an allocated Fog node}
   $m=m-1$\;
   Remove an allocated Fog node\;
   Calculate $\overline{W_1}(m)$\;}
  }
  Calculate $\overline{W_1}(m)$ and $\overline{W_2}$\;
  \eIf{$W^{th}_{SCT} > \overline{W_2}$}{
   System Critical Tasks $\in$ class 2\;
   $\overline{W_{SCT}} = \overline{W_2}$\;
   }{
   System Critical Tasks $\in$ class 1\;
   Calculate $\rho_1$ and $\rho_2$ using \eqref{newrho}\;
   $\overline{W_{SCT}} = \overline{W_1}(m)$\;
  }
  Calculate $\rho_1$, $\rho_2$ and $\rho$\;
 }
 \caption{Algorithm for dynamic resource allocation at FCL.}
 \label{algo1}
\end{algorithm}

In algorithm \ref{algo1}, we present our proposed algorithm for priority-aware task scheduling and dynamic resource allocation. We start by initializing all the parameters followed by calculation of the server utilization and waiting time for the respective classes. If the waiting time of the second class exceeds the threshold for SCTs, then the SCTs will be reallocated to class 1 else they remain in class 2. Followed by this we recalculate the respective waiting times and check whether the waiting time for class 1 exceeds the class 1 threshold. If yes, then we add fog nodes into the FCL until the waiting time comes below the threshold or the number of fog nodes becomes equal to the maximum allowed or available fog nodes. If no, then we check for whether there are an excessive number of fog nodes allocated, if so, then we remove some fog nodes in such a way that the waiting time comes near the threshold waiting time but also considering that it does not deem the system unstable.

\section{Numerical results}\label{results}
In this section we evaluate our proposed algorithm with other schemes and present its results. We also evaluate the efficiency of our algorithm by using a power consumption parameter.

\subsection{Simulation setup}
We use the analytical model given in Section \ref{promodel} and implement our proposed algorithm in Python. We also use Java Modelling Tool (JMT) which is a free open-source tool for the performance evaluation of queueing models to verify our results. The parameters used during the simulation are clearly mentioned in Table \ref{simpar}.

\begin{table}
 \caption{Simulation Parameters}
\label{simpar}
\begin{tabular}{|p{1.5cm}|p{4cm}|p{1.4cm}|}
\hline
\textbf{Parameters}     & \textbf{Description} & \textbf{Value}  \\ \hline
\hline
$\lambda$ & Task arrival rate & 1-14 tasks/s  \\ \hline
$\mu$ & Service rate per server & 1 tasks/s \\ \hline
$\alpha$  & Ratio of delay-sensitive tasks to all tasks & 0.2 \\ \hline
$\beta$       & Ratio of system critical tasks to all tasks & 0.1  \\ \hline
$m$        & Initial number of fog nodes & 18   \\ \hline
$m_{max}$      & Maximum number of fog nodes & 20   \\ \hline
$m_{min}$      & Minimum number of fog nodes & 15   \\ \hline
$W_1^{th}$       & Delay threshold for class 1 tasks & 0.01 sec \\ \hline
$W_{SCT}^{th}$        & Delay threshold for SCT & 0.02 sec \\ \hline
\end{tabular}
\centering
\end{table}

\subsection{Results}
In this subsection we present the numerical results of our algorithm.

\begin{figure}[t]
\includegraphics[width=\linewidth]{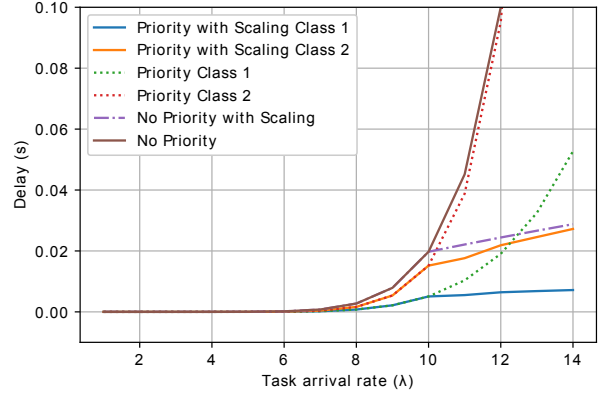}
\centering
\caption{Plot for delay (sec) vs rate of arrival of tasks.}
\label{fig:delay}
\end{figure}

In Fig. \ref{fig:delay}, the waiting time for the proposed algorithm is compared with other schemes such as only priority scheme, no priority but with scaling scheme and no priority and no scaling scheme. It can be seen from the plot that the proposed algorithm performs better than any other scheme as the delay observed at higher values of $\lambda$ is least in the proposed scheme whereas for schemes without scaling the delay increases exponentially. Delay for delay-sensitive tasks for the proposed approach remains below the provided threshold which is 0.01 seconds and the delay for the delay-insensitive tasks also remains significantly low. 

\begin{figure}[t]
\includegraphics[width=\linewidth]{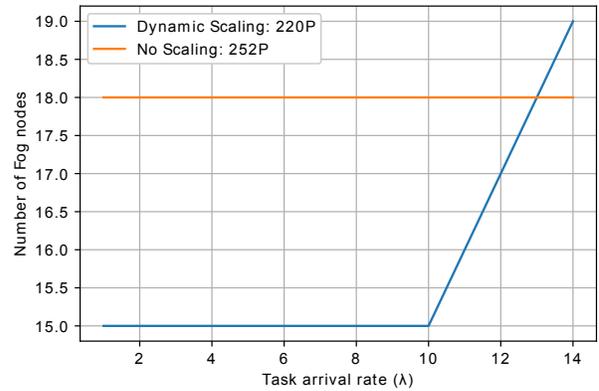}
\centering
\caption{Plot for number of servers at FCL for both dynamic and no scaling schemes with increasing rate of arrival of tasks.}
\label{fig:pow}
\end{figure}

From the Fig. \ref{fig:delay}, it can be argued that since the number of fog nodes increase to reduce the delay, it may result in higher power consumption. But in Fig. \ref{fig:pow}, we show that the net power consumed by the fog nodes for the proposed scheme is lower than the power consumed by non-scalable scheme. The initialized number of fog nodes for the simulation is provided as 18. In case of non-scalable schemes the number of fog nodes remain the same which means for rate of arrival of tasks, more than required number of fog nodes are allocated. Furthermore, as the rate of arrival of tasks increases, the allocated fog nodes are not able to cope up with the tasks, resulting in significant increase in processing time. Whereas in the proposed scheme, the number of fog nodes are allocated between a defined value of minimum and maximum fog nodes. For the simulations we have taken $m_{min}=15$ and $m_{max}=20$. Thus, initially when the rate of tasks arrival is low, the scheme switches to a lower number of fog nodes and as the rate increases, more and more nodes are allocated in the FCL. We assumed that each fog nodes consumes $P$ units of power for completing the tasks at each instance. Thus, we can see that for non-scalable scheme the power consumed is $252P$, whereas for the proposed scheme, the power consumed is $220P$, which is $14.5\%$ lower. Therefore, it is evident that the proposed approach is able to achieve a significantly lower delay with an even lower power consumption. This plot justifies the proposed scheme to be superior to other schemes.

\begin{figure}[t]
\includegraphics[width=\linewidth]{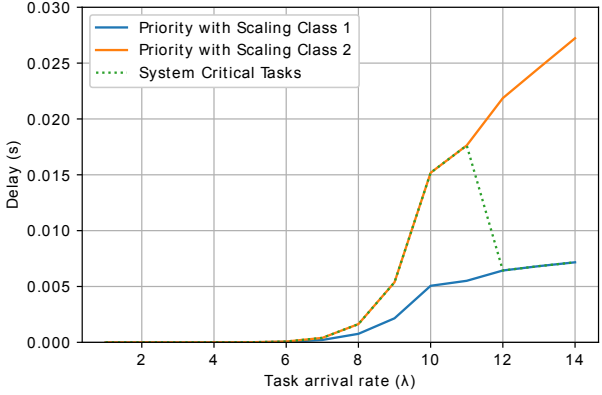}
\centering
\caption{Plot showcasing the class switching of SCT with increasing rate of arrival of tasks for both dynamic and no scaling schemes.}
\label{fig:sct}
\end{figure}

Figure \ref{fig:sct} shows the class switching of SCT tasks, as the delay of class 2 tasks is about to cross the threshold delay for SCT which is 0.02 seconds. The System Critical Tasks is allocated along with class 1.

\begin{figure}[t]
     \centering
     \begin{subfigure}[b]{0.49\linewidth}
         \centering
         \includegraphics[width=\linewidth]{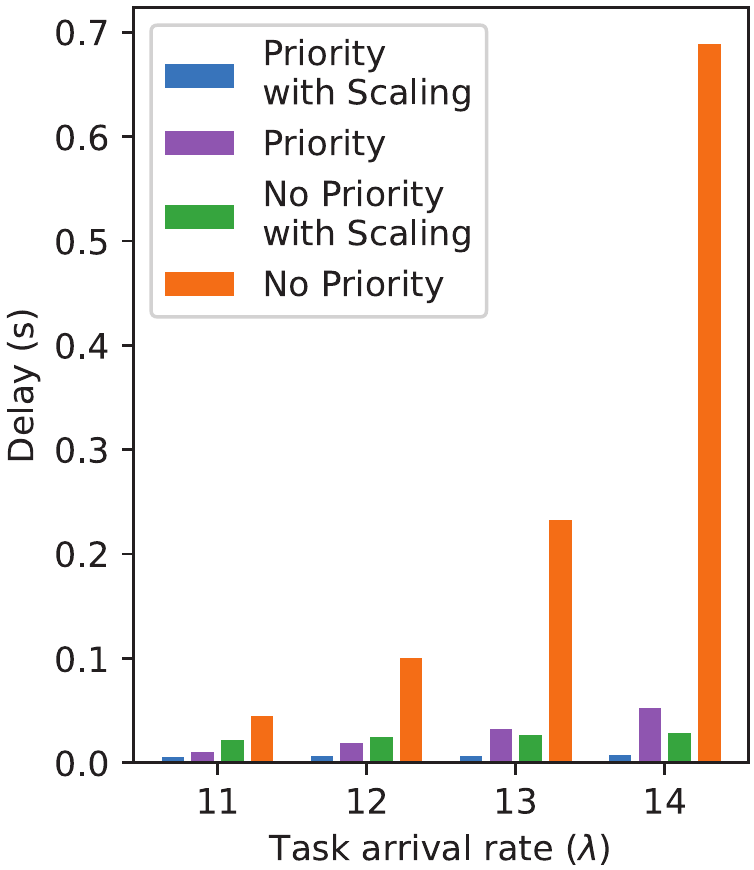}
         \caption{Delay-sensitive tasks}
         \label{fig:bardel1}
     \end{subfigure}
     \hfill
     \begin{subfigure}[b]{0.49\linewidth}
         \centering
         \includegraphics[width=\linewidth]{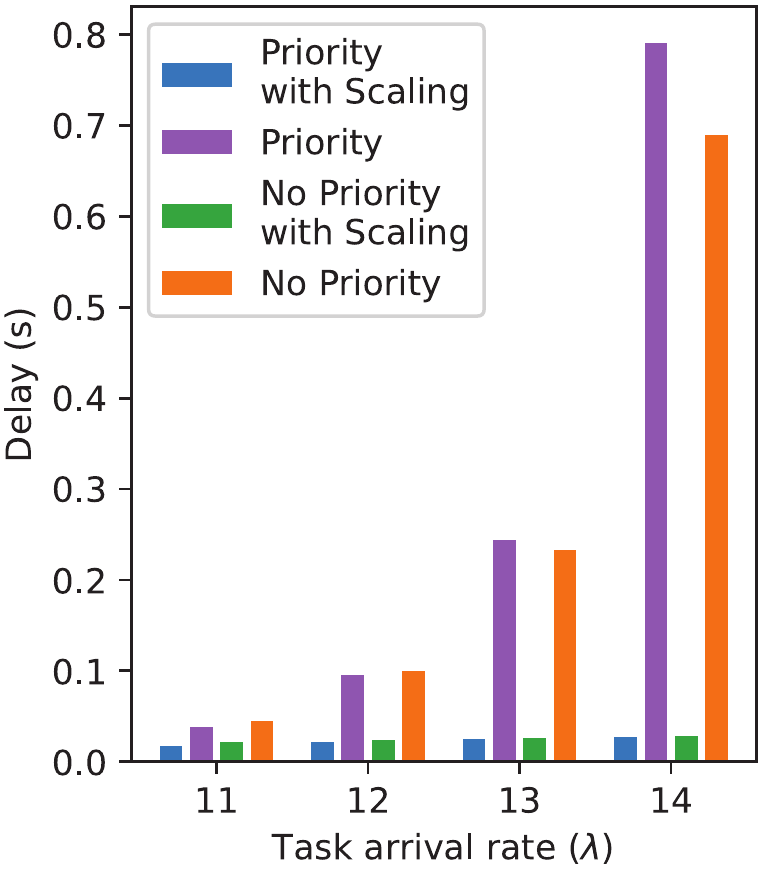}
         \caption{Delay-insensitive tasks}
         \label{fig:bardel2}
     \end{subfigure}
        \caption{Plot for Delay (s) at different rate of arrival of tasks ($\lambda$).}
        \label{fig:bardel}
\end{figure}

% \begin{figure}[t]
% \includegraphics[width=\linewidth]{serverutil.eps}
% \centering
% \caption{Plot for server utilization at the FCL vs rate of arrival of tasks.}
% \label{fig:server}
% \end{figure}

% Figure \ref{fig:server} shows the plot for server utilization for scalable as well as non-scalable schemes. As the number of fog nodes increases the server utilization decreases in the proposed scalable scheme, which shows that the fog nodes will not be completely loaded as the arrival rate of the tasks increases.

From the numerical results, we can conclude that the proposed scheme is able to achieve significantly lower delay for both the classes as well as it is being achieved with more efficient use of power.

\section{Conclusion}\label{conclusion}
A priority based queueing scheme with scalable fog servers was proposed in this paper for tackling the two major research problems in Fog computing paradigm simultaneously, which is efficient task scheduling and resource allocation. The queues were modelled as $M/M/m$ queues with delay-sensitive tasks as higher priority tasks and delay-insensitive tasks as lower priority tasks. Furthermore, a scalability scheme was also combined which allowed dynamic allocation of fog servers depending upon the computational load on the system. Upon performance evaluation of the proposed model we showcased that the proposed algorithm was able to achieve significantly lower delay for both delay-sensitive and -insensitive tasks with $14.5\%$ lower power consumption.
% \section*{Acknowledgment}
% We are looking for agencies which can fund the submission of this paper in a reputed conference.
%\section*{References}
\bibliographystyle{IEEEtran}
\bibliography{main}

\end{document}